\documentclass[prl,twocolumn,showpacs,preprintnumbers,amsmath,amssymb]{revtex4}
\usepackage{slashed}

\begin{document}

\preprint{IZTECH-P2011/05}

\preprint{\bf arXiv:1110.3815 [hep-ph]}

\title{Gravity wraps Higgs boson}

\author{Durmu\c{s} A. Demir}
\affiliation{Department of Physics, Izmir Institute of
Technology, IZTECH, TR35430 Izmir, Turkey}

\date{\today}

\begin{abstract}
It is shown that, under a conformal transformation with reference to the
Higgs field, the Higgs boson can be completely decoupled from
electroweak interactions with no apparent change in known properties of leptons, quarks
and vector bosons. Higgs boson becomes part of a scalar-tensor gravity
which can be relevant for Dark Energy. It interacts with matter
sector via higher-dimensional terms ({\it e.g.} neutrino Majorana mass), and via the
fields (of new physics) whose masses are not generated by the Higgs mechanism. Dark Matter
and two-Higgs-doublet model are the simplest examples.

\end{abstract}

\maketitle

The electroweak theory  is a  isospin $SU(2)_L$ times hypercharge $U(1)_Y$
gauge theory \cite{weinberg,salam,glashow}
spontaneously broken down to electric charge $U(1)_{e}$ via Higgs
mechanism
\cite{higgs-mech,aberslee}. Higgs field, the order parameter of the
electroweak transition, transforms as
$\left[{\bf 2}, \frac{1}{2}\right]$ under $\left[SU(2)_L, U(1)_Y\right]$,
\begin{eqnarray}
\label{higgs-gen}
H = \frac{\varphi}{\sqrt{2}}\, e^{\frac{i}{\upsilon}
\overrightarrow{{\phi}}\cdot \overrightarrow{T} }
\left(\begin{array}{c}0\\ \upsilon\end{array}\right)
\end{eqnarray}
where $\overrightarrow{T}$ are the three $SU(2)_L$ generators weighted by
the Goldstone boson fields $\overrightarrow{\phi}$, and
\begin{eqnarray}
\label{var-phi}
\varphi = 1 + \frac{h}{\upsilon}
\end{eqnarray}
is the modulus field encoding the Higgs boson $h$.  The constant
$\upsilon$ is the vacuum expectation value (VEV) of the neutral component
of the Higgs doublet.

Gauge invariance ensures that, it is always possible to perform an
$SU(2)_L$ rotation to  weed out
the Goldstone bosons in (\ref{higgs-gen}) so that Higgs field gets mapped
into the representation in the unitary gauge
\begin{eqnarray}
\label{unitary}
H_U = \frac{\varphi}{\sqrt{2}} \left(\begin{array}{c}0\\
\upsilon\end{array}\right)
\end{eqnarray}
everywhere in the electroweak Lagrangian \cite{aberslee}.  The Higgs boson
$h$ is the remnant, and there can exist no gauge transformation that can
erase it:  It is in the spectrum as a real scalar particle, and it couples
to all the particles whose masses are generated by the Higgs mechanism.

Inspired by the role of the gauge transformations in removing the
Goldstone bosons, one wonders whether it is possible to remove the Higgs
boson itself from electroweak interactions partially, if not entirely.
This will be shown to be possible via the scalings of the fields by
appropriate powers of the modulus $\varphi$. In particular, as was first
proposed in \cite{demir}, it will be proven that the Higgs boson can be
wholly transferred from the matter sector to the gravity sector where it
possesses only gravitational interactions and  escapes detection in
experiment. Though Higgs boson is completely decoupled from the
electroweak sector, the leptons, quarks and vector bosons  possess their
already known properties: The observed masses, couplings, and mixings.
However, there is no Higgs boson around to interact with. The Higgs boson
gives rise to a scalar-tensor theory of gravity where strength of the
gravitational interactions stays put at Newton's constant (provided that
Higgs boson stays perturbative \cite{demir}). For realizing these novel
features, it is beneficial to introduce
\begin{eqnarray}
\label{trans-Higgs}
H_U = \varphi^{a_0} {\mathbb{H}}
\end{eqnarray}
as a transformation rule which peals off $\varphi$ from $H_{U}$ to reduce
it to
\begin{eqnarray}
\label{vev}
{\mathbb{H}} = \frac{1}{\sqrt{2}} \left(\begin{array}{c} 0 \\
\upsilon\end{array}\right)
\end{eqnarray}
with $a_0=1$.

The relation (\ref{trans-Higgs}) hints at a similar scaling transformation
for fermions
\begin{eqnarray}
\label{trans-fermion}
F = \varphi^{a_{1/2}} {\mathbb{F}}
\end{eqnarray}
where $F$ denotes collectively the leptons ( $L$ $\left[{\bf 2},
\frac{1}{2}\right]$, $E$ $\left[{\bf 1},  -{1}\right]$) and quarks ( $Q$
$\left[{\bf 2}, \frac{1}{6}\right]$, $U$ $\left[{\bf 1},
\frac{2}{3}\right]$, $D$ $\left[{\bf 1}, -\frac{1}{3}\right]$).

The $U(1)_Y$ gauge field $Y_{\alpha}$ $\left[{\bf 1}, 0\right]$ and
$SU(2)_L$ gauge field ${\overrightarrow{W}}_{\alpha}$ $\left[{\bf 3},
{0}\right]$ must be invariant
\begin{eqnarray}
\label{trans-gauge}
Y_{\alpha} = {\mathbb{Y}}_{\alpha}\,,\; \overrightarrow{W}_{\alpha} =
{\overrightarrow{\mathbb{W}}}_{\alpha}
\end{eqnarray}
as they pertain to gauge transformations.

The transformations (\ref{trans-Higgs}), (\ref{trans-fermion}) and
(\ref{trans-gauge})  are realized differently by different sectors of
the electroweak theory \cite{bekenstein}. For instance, the kinetic terms
of the fermions
\begin{eqnarray}
\label{fermions}
- \frac{1}{2} \sum_{F=L, Q, E, U, D} {\overline{F}}
\left({\overrightarrow{\slashed{\cal{D}}}} -
{\overleftarrow{\slashed{\cal{D}}}} \right) F
\end{eqnarray}
scale as $\varphi^{2 a_{1/2}}$. The Yukawa interactions
\begin{eqnarray}
\label{yukawa}
-  \left[ h_E \overline{L} H E + h_D \overline{Q} H D + h_U \overline{Q}
H^{c} U  + {\mbox{H. C.}}\right]
\end{eqnarray}
behave $\varphi^{2 a_{1/2} + a_0}$. The kinetic terms of the gauge fields
\begin{eqnarray}
\label{gauge bosons}
-\frac{1}{4} {{g}}^{\mu\alpha} {{g}}^{\nu\beta} Y_{\mu\nu} Y_{\alpha\beta}
- \frac{1}{4}  {{g}}^{\mu\alpha} {{g}}^{\nu\beta} {\overrightarrow{
W}}_{\mu\nu} \cdot {\overrightarrow{ W}}_{\alpha\beta}
\end{eqnarray}
are obviously invariant due to (\ref{trans-gauge}). In contrast,  the
Higgs sector
\begin{eqnarray}
\label{higgs-sect}
- {{g}}^{\alpha\beta} \left({\cal{D}}_{\alpha} H\right)^{\dagger}
\left({\cal{D}}_{\beta} H\right) - m_H^2 H^{\dagger} H - \lambda \left(
H^{\dagger} H\right)^2
\end{eqnarray}
behaves rather heterogeneously, since each term involves different powers
of $\varphi$.

The scaling characteristics of (\ref{fermions}), (\ref{yukawa}),
(\ref{gauge bosons}) and (\ref{higgs-sect}) do not accommodate an
invariance principle. Indeed, imposing invariance directly gives $a_{1/2}
= 0$ and $a_{0} = 0$, which are trivial. Therefore,  for achieving
invariance, even at a partial level, an extra agent is needed; the metric
tensor $g_{\alpha\beta}$. Namely, $g_{\alpha\beta}$ must be elevated to
the status of a  dynamical variable. It is to scale as
\cite{salam2,demir}
\begin{eqnarray}
\label{trans-metric}
g_{\alpha\beta} = \varphi^{a_2} {\mathbb{G}}_{\alpha\beta}
\end{eqnarray}
 concurrently with (\ref{trans-Higgs}), (\ref{trans-fermion}) and
(\ref{trans-gauge}). Having metric introduced in the game, the total
action becomes
\begin{eqnarray}
\label{action}
S_{tot} =  \int d^{4}x \sqrt{ - g}\, \left[{\mathfrak{L}}_{GR} - \lambda_R
{\cal R}\, H^{\dagger} H + {\mathfrak{L}}_{EW} \right]
\end{eqnarray}
where, in the minimal case (not including higher-curvature terms or fields
beside the metric tensor), its gravitational part
\begin{eqnarray}
\label{gr-lagran}
{\mathfrak{L}}_{GR} = \frac{1}{2} M_{G}^2 \left( {\cal R} - \Lambda
\right)
\end{eqnarray}
is the Einstein-Hilbert term with cosmological constant $\Lambda$ and bare
gravitational scale  $M_{G}$ such that
\begin{eqnarray}
\overline{M}_{Pl}^2 = M_G^2 - \frac{1}{2} \lambda_R \upsilon^2
\end{eqnarray}
is the Planck scale induced by the non-minimal coupling in (\ref{action}).

Part of the action (\ref{action}) explicitly involves the Higgs boson
\cite{aberslee}
\begin{eqnarray}
\label{higgs-dep-lagran}
{\mathfrak{L}}(h) &=& - \frac{1}{2} \upsilon^2 \left[
g^{\alpha\beta}\partial_{\alpha}\varphi \partial_{\beta}\varphi +  m_{H}^2
\varphi^2 + \frac{\lambda}{2} \upsilon^2 \varphi^4\right] \nonumber\\
&-& \frac{1}{2} \varphi^2 g^{\alpha\beta} \left[ M_W^2 W^{+}_{\alpha}
W^{-}_{\beta} + M_{Z}^2 Z_{\alpha} Z_{\beta}\right]\nonumber\\
&+& \varphi \sum_{F=\ell, u, d} m_{F} \overline{F} F + \frac{1}{2}
\lambda_R \varphi^2 {\cal R}
 \end{eqnarray}
while the rest is entirely free from the Higgs boson. Expanding this in
powers of $h$, at the lowest order, the vacuum energy becomes
\begin{eqnarray}
\label{min-en}
V(\upsilon) = M_{G}^2 \Lambda + \frac{1}{2} m_H^2 \upsilon^2 +
\frac{\lambda}{4} \upsilon^4
\end{eqnarray}
if the gravitational sector deposits only the cosmological term $M_{G}^2
\Lambda$. The vacuum energy density $V(\upsilon)$ is minimized for
\begin{eqnarray}
\label{vev-matt}
\upsilon^2 = - \frac{m_H^2}{\lambda}
\end{eqnarray}
which makes sense if Higgs is tachyonic {\it i. e.} $m_H^2 < 0$. En
passant, one notices that $V(\upsilon)$ can be tuned to vanish by taking
\begin{eqnarray}
\label{cc-matt}
\Lambda = \frac{\lambda}{4}  \left(\frac{v^2}{M_{G}}\right)^{2}
\end{eqnarray}
with $\upsilon$ is given by (\ref{vev-matt}). With this tuning, because of
vanishing $V(\upsilon)$, the background metric $g^{(0)}_{\alpha\beta}$
(which must nullify the Ricci scalar because of vanishing $V(\upsilon)$)
can be identified with  the flat Minkowski metric $\eta_{\alpha\beta}$.

After transforming the action (\ref{action}) according to the rules in
equations (\ref{trans-Higgs}), (\ref{trans-fermion}), (\ref{trans-gauge})
and (\ref{trans-metric}), one finds that part of the electroweak
Lagrangian ${\mathfrak{L}}_{EW}$ becomes independent of $\varphi$ while
the gravity sector ${\mathfrak{L}}_{GR}$ metamorphoses into a
scalar-tensor theory by swallowing $\varphi$ \cite{salam2,demir}. This
happens for the specific values of the exponents
\begin{eqnarray}
a_{1/2} = \frac{3}{2}\,,\; a_2 = - 2
\end{eqnarray}
which are, together with $a_0 = 1$, nothing but the conformal weights of
the fields arising in scale transformations \cite{bekenstein,demir}. With
these weights, the image of (\ref{action}) becomes
\begin{eqnarray}
\label{action-bar}
{\mathbb{S}}_{tot} =  \int d^{4}x \sqrt{ - {\mathbb{G}}}\,
\left[{\mathfrak{L}}_{EW}(h=0) + {\mathfrak{L}}_{ST}({\mathbb{G}},h)
\right]
\end{eqnarray}
wherein the first term is nothing but the usual electroweak Lagrangian
${\mathfrak{L}}_{EW}$ with Higgs boson $h$ set to zero everywhere, and the
second term refers to a scalar-tensor theory of gravity which involves
only gravity and the Higgs boson:
\begin{eqnarray}
\label{higgs-dep-lagran2}
{\mathfrak{L}}_{ST} &=& \frac{1}{2} \left[ 6 M_{G}^2 \varphi^{-4} - \left(
1 - 6 \lambda_R\right) \upsilon^2 \varphi^{-2}\right]
{\mathbb{G}}^{\alpha\beta} \partial_{\alpha}\varphi
\partial_{\beta}\varphi\nonumber\\
&-& \frac{1}{4} \lambda \upsilon^4 - \frac{1}{2} m_{H}^2 \upsilon^2
\varphi^{-2} - M_{G}^2 \Lambda \varphi^{-4}\nonumber\\
&+& \frac{1}{2}\left[ {M}_{G}^2 \varphi^{-2}
- \lambda_R \upsilon^2\right] {{\mathbb{R}}}
\end{eqnarray}
where ${\mathbb{R}}\left({\mathbb{G}}\right)$ is the image of ${\cal
R}\left({g}\right)$ under (\ref{trans-metric}).

Comparison of the two Lagrangians, (\ref{higgs-dep-lagran}) and
(\ref{higgs-dep-lagran2}), manifestly shows that the  Higgs boson $h$ is
completely transferred from the matter sector to the gravity sector.  The
Higgs boson interactions with gauge bosons and fermions in
(\ref{higgs-dep-lagran}) convert into pure mass terms.

For revealing the physics implications of the scalar-tensor gravity
(\ref{higgs-dep-lagran2}), it is convenient to expand (\ref{action-bar})
about the vacuum configuration $\varphi=1$ to find
\begin{eqnarray}
\label{higgs-dep-lagran3}
{\mathfrak{L}}_{ST} &=& - \frac{1}{2} \left[ \delta^2 - \lambda_1 P(h) -
\lambda_2 P(h)^2 \right] {\mathbb{G}}^{\alpha\beta}
\left(\partial_{\alpha} h\right) \left(\partial_{\beta}
h\right)\nonumber\\
&-& \frac{1}{2} \left[ m_{H}^2 \upsilon^2 + 4 M_{G}^2 \Lambda\right] P(h)
- M_{G}^2 \Lambda P(h)^2\nonumber\\
&+& \frac{1}{2} \left[ {\overline{M}}_{Pl}^2 + M_{G}^2 P(h)\right]
{\mathbb{R}} - V(\upsilon)
\end{eqnarray}
where $\lambda_1 = \lambda_2 - \delta^2 = 1- 6 \lambda_R - 2 \delta^2$
with
\begin{eqnarray}
\delta^2 = 1 - 6 \lambda_R - \frac{ 6 M_{G}^2}{\upsilon^2}
\end{eqnarray}
which must be positive and nonvanishing for Higgs boson not to be a ghost.
The function
\begin{eqnarray}
P(h) = \sum_{n=1} (-1)^n (n+1) \left(\frac{h}{\upsilon}\right)^n
\end{eqnarray}
results from the expansion of $\varphi^{-2}$. The vacuum energy is
$V(\upsilon)$ as in (\ref{min-en}), and it vanishes upon (\ref{vev-matt})
and (\ref{cc-matt}) to enable identification of
${\mathbb{G}}^{(0)}_{\alpha\beta}$ with $\eta_{\alpha\beta}$. In fact,
$g^{(0)}_{\alpha\beta}$ and  ${\mathbb{G}}^{(0)}_{\alpha\beta}$ are
expected to be identical from (\ref{trans-metric}). Unlike its finite
polynomial interactions in (\ref{higgs-dep-lagran}), the Higgs boson
develops now infinite-order interactions in (\ref{higgs-dep-lagran3}) yet
its mass-squared, $m_{h}^2 = - 2 m_{H}^2$, stays put at its value in the
electroweak theory. These features ensure that the vacuum structure of the
theory does not change in passing from (\ref{higgs-dep-lagran}) to
(\ref{higgs-dep-lagran2}).

Having structured and audited the mechanism, it could be enlightening to
highlight and discuss its certain salient features. This is done below.
\begin{enumerate}

\item {\it Frames for Higgs Boson}. Collider experiments of decades have
determined almost all of the parameters of the
electroweak theory. The only yet-to-be-observed piece is the Higgs boson
$h$ \cite{cavidi}. The analyses of
the Tevatron and LHC data are continuing, and it is likely that Higgs
boson will be
found to have a mass within one of those narrow intervals indicated by the
most
recent searches \cite{experiment}. It is, however, also likely that the
Higgs boson
will not be discovered at all.

The main outcome of the mechanism is that, particle colliders may not
discover Higgs boson
not because the electroweak theory is not working but because gravity
secludes the Higgs boson from
leptons, quarks and vector bosons. Indeed, gravity is there to couple
everything, and conformal
transformations (\ref{trans-Higgs}), (\ref{trans-fermion}),
(\ref{trans-gauge}) and (\ref{trans-metric})
switch the interaction scheme invariably from (\ref{action}) (to be called
{\it Standard frame} to mean the
usual Einstein frame) to (\ref{action-bar}) (to be called {\it Gravic
frame} to mean the usual Jordan frame ). These two frames, as was first
noted in \cite{demir}, give strikingly opposite predictions for Higgs
boson. In particular, colliders cannot access the Higgs boson in the
Gravic frame due to its secluded nature not due to weakening of its signal
by light singlet scalars \cite{singlets}.

\item {\it Renormalizability and Unitarity.} Electroweak theory is
renormalizable in Standard frame in the absence of gravity.  In fact, it
is gravity which renders theory nonrenormalizable in the Standard frame,
it is gravity which facilitates the Gravic frame, and hence, it must
wholly be gravity which causes nonrenormalizablity in the Gravic frame.

    Scattering amplitudes of longitudinal weak bosons grow as $g^2
E^2/M_W^2$ with the center-of-mass energy $E$. Thus, unitarity is
violated at the scale $E_{crit} \simeq 1.2\ {\rm TeV}$. In Standard frame,
Higgs boson restores unitarity if $m_h \lesssim 1\ {\rm TeV}$
\cite{quigg}. In Gravic frame, unitarity is lost at $E_{crit}$ yet Higgs
boson couples to gravity and fields whose masses are not generated by the
Higgs mechanism ( supersymmetric partners, Kaluza-Klein levels or
technicolor fields which can exist in the ${\rm TeV}$ domain).

\item {\it Higher-Dimensional Operators}. These are best exemplified by
massive neutrinos. A Dirac neutrino acquires mass via the Yukawa
interaction
\begin{eqnarray}
{\mathfrak{L}}_{EW} \ni - h^{D}_{\nu} \overline{L} H^{c} \nu_R + {\mbox{H.
C.}}
\end{eqnarray}
which, under the scalings (\ref{trans-Higgs}), (\ref{trans-fermion}) and
(\ref{trans-metric}) from the Standard frame (\ref{action}), reduces to
a pure neutrino mass term
\begin{eqnarray}
\label{mass-dir}
{h^{D}_{\nu}} \frac{\upsilon}{\sqrt{2}} \overline{\nu_L} \nu_R + \mbox{H.
C.}
\end{eqnarray}
in the Gravic frame of (\ref{action-bar}). Therefore, Dirac neutrinos,
like all charged leptons and quarks,
exhibit no interactions with the Higgs boson in the Gravic frame.

A Majorana neutrino, on the other hand, acquires mass from the seesaw term
\begin{eqnarray}
{\mathfrak{L}}_{EW} \ni - \frac{1}{M_R} h^{M}_{\nu} \left(\overline{L}
H^{c}\right) \left( H^{T} L^c\right) + \mbox{H. C.}
\end{eqnarray}
which, under the scalings (\ref{trans-Higgs}), (\ref{trans-fermion}) and
(\ref{trans-metric}) from the Standard frame (\ref{action}), becomes
\begin{eqnarray}
\label{mass-maj}
h^{M}_{\nu} \frac{\upsilon^2}{2 M_R} \varphi \overline{\nu_L} \nu_{L}^c  +
\mbox{H. C.}
\end{eqnarray}
in the Gravic frame of (\ref{action-bar}). Therefore, a Majorana neutrino
interacts with the Higgs boson in both Standard and Gravic frames. The two
neutrino types are strikingly different. This is because Majorana
neutrinos get their masses from a  dimension-5 operator.

Higgs boson contributions to higher-dimensional operators cannot be
pared off due to the presence of a mass scale. Therefore, Higgs boson can
interact with all the matter species even in the Gravic frame in case
Standard frame involves appropriate higher-dimensional operators.

Higher-dimensional operators are generated also by quantum gravitational
effects. Indeed, small perturbations about the vacuum configuration (about
$g^{(0)}_{\alpha\beta}$ or ${\mathbb{G}}^{(0)}_{\alpha\beta}$ in the two
frames) form gravitational waves, whose quantization, if can ever be done,
creates loops of gravitons. Thus,  in the Gravic frame, the Higgs boson
and matter sector, which communicate with each other through the metric
tensor at the tree level, develop Planck-suppressed higher-dimensional
direct interactions via graviton exchange.

\item {\it Astrophysical and Cosmological Effects}. The Higgs boson is
unique in that it is  the only massive, spinless member of the electroweak
spectrum. In Standard frame, it mediates interactions between scalar
currents. Furthermore, it is known to give successful chaotic inflation
where inflation starts at Planckian values of $h$ and ends at the
electroweak vacuum $h=0$ \cite{higgs-inf}. This scenario ascribes a novel
role  and epithet to Higgs boson with certain difficulties \cite{problems}
and associated cures \cite{bauer}. The Higgs boson Lagrangian
(\ref{higgs-dep-lagran}), together with the non-minimal coupling term in
(\ref{action}), is precisely the one used in \cite{higgs-inf}.

    In the Gravic frame, excepting higher-dimensional operators, the Higgs
boson interacts only with gravity. This immunity of the Higgs boson to all
forces but gravity is precisely what is required of the models of Dark
Energy. The reason is that, Dark Energy, an example of which is the vacuum
energy, participates only in  gravitational interactions \cite{model}. It
is thus highly likely that Higgs boson in Gravic frame pertains to Dark
Energy, and it may cause an accelerated expansion for the Universe. This
is expected on the basis of its Lagrangian (\ref{higgs-dep-lagran2}) which
generalizes the non-minimally coupled scalar field theories which are
known to yield accelerated expansion  \cite{faraoni}. Besides this, the
non-standard Majorana neutrino coupling to Higgs boson in (\ref{mass-maj})
can model the suggested role \cite{neutri} of neutrinos in forming Dark
Energy.

    The Higgs boson in Gravic frame, as it stands in
(\ref{higgs-dep-lagran2}) or (\ref{higgs-dep-lagran3}), interacts only
with gravity. It is thus immune to various bounds from solar system and
other astrophysical structures \cite{astro1}. Even if it interacts, via
higher-dimensional operators, the Gravic  Lagrangian
(\ref{higgs-dep-lagran2}) can be put in the Brans-Dicke form with
${\phi}_{BD}= (M_G^2 \varphi^2 - \lambda_R
\upsilon^2)/\overline{M}_{Pl}^2$ which is too massive (around neutrino
mass) to have observable effects on the solar system \cite{astro2}. This
can also be seen by reverting to the Standard frame.

    One also notes that, instead of the minimal gravitational sector in
(\ref{gr-lagran}), it is possible to consider  an  $F({\cal{R}})$ gravity
or a dilatonic gravity or a more general structure. The essence of the
mechanism does not change.

\end{enumerate}

The present work has thus established that, the Higgs boson can be completely
decoupled from electroweak interactions with no changes in known properties
of quarks, leptons and vector bosons. Higgs boson becomes part of a scalar-tensor gravity
which can be relevant for Dark Energy. In this frame, Higgs boson does not couple to
fields whose masses are generated by the Higgs mechanism. It couples to
matter sector via higher-dimensional terms or via the fields of new physics
(related to  Dark Matter,  multi-Higgs-doublet models, supersymmetry, extra dimensions or technicolor).

The author is grateful to Bar{\i}{\c s} Ate{\c s}, Canan Karahan and Hale
Sert for fruitful discussions. He  is thankful to Nam{\i}k K. Pak for his
suggestions and careful reading of the manuscript.

\end{document}